\documentstyle[prl,aps,epsf]{revtex}

\begin{document}
\draft

\twocolumn[\hsize\textwidth\columnwidth\hsize\csname 
@twocolumnfalse\endcsname

\title{Liquid Surface Wave Band Structure
Instabilities}

\author{Tom Chou}

\address{DAMTP, University of Cambridge, Cambridge
CB3 9EW, ENGLAND}

\date{\today}
\maketitle

\begin{abstract}

We use Bloch wavefunctions in a new way to study a {\it fluid}
interfacial problem and find both linear oscillatory and nonoscillatory
instabilities of the surface deformation. Underlying periodic flows,
such as those arising in Rayleigh-B\'{e}nard cells, Langmuir
circulation, and solar convection rolls are treated as regions of
varying surface shear which scatter and refract surface
capillary-gravity waves. We find that Bloch wavefunction decompositions
of the surface deformation $\eta(\vec{r})$ and velocity potential
$\varphi(\vec{r},z)$ results in a nonhermitian secular matrix with an
associated band structure that gives rise to extremely rich surface
instabilities (complex frequencies).  These are predominantly enhanced
at certain Brillouin zone edges, particularly near Bragg planes
corresponding to the periodicity determined by converging or diverging
surface flows. The instabilities persist {\it even} when the dynamical
effects of the upper fluid are neglected, in contrast to the uniform
shear Kelvin-Helmholtz instability.  The periodic flows can also couple
with uniform shear and {\it suppress} standard Kelvin-Helmholtz(K-H)
instabilities.  

%With uniform shear,
%the surface wave band structure also displays an ``optical'' branch
%with a ``massive'' zero wavevector mode gap, in analogy with phonons
%of a Bravais lattice with an internal basis.  

\end{abstract}
\pacs{47.20.Ft, 47.20.Ma, 68.10.-m}
]

%\newpage

{\it Introduction - } Regular fluid flow patterns are ubiquitous
\cite{CRAIK}. Periodic flows are typically the result of thermal and/or
dynamical instabilities and can exist in parameter regimes prior to
chaotic or turbulent flows.  An example on laboratory scales is the
well-known Rayleigh-B\'{e}nard convective instability \cite{LEAL}. 
Heating of a fluid layer from below provides buoyancy which induces
the instability.  As the instability develops, regular convective
patterns in the form of rectangles and hexagons appear
\cite{LEAL,JOSEPH}.  

Another physical example of nearly periodic flows with a free surface occurs in
Langmuir circulation (LC) windrows where wind stresses, turbulent stresses, and
coriolis forces conspire to form convection rolls in the upper ocean
\cite{LEIB,LC}. These flows can be spatially periodic as often observed when
the sea surface is contaminated by oil slicks or algae.  The LC process is
believed to be an important mechanism in thermal and chemical mixing in the
upper ocean. Mechanisms of LC generation may involve the interaction of
turbulent surface currents with surface waves which in turn mediate the wind
stresses \cite{LC}. Refraction of largely irrotational surface waves from the
underlying rotational Langmuir circulation rolls are thought to contribute to
wave breaking and enhanced local wind stresses \cite{LC}. On an even larger
scale are the approximately periodic solar convection cells in the solar
convective zone.  In this system, magnetic fields will likely also affect the
surface wave dynamics \cite{SHERCLIFF}.

Clearly, surface wave propagation and interfacial stability is an
important aspect of stratified fluids with wide applicability. 
Oscillatory instabilities of the two-layer B\'{e}nard problem have been
studied by considering all coupled modes, including thermal and Marangoni
\cite{JOSEPH}.  In the limiting one layer Rayleigh-B\'{e}nard problem,
Benguria and Depassier \cite{BENGURIA} even find interfacial oscillatory
instabilities for parameters {\it prior} to the onset of the periodic
roll states. This occurs in the fixed lower temperature and fixed upper
thermal flux ensemble. In this Letter, we assume a pre-existing periodic
flow and explore its dynamical effects on interfacial instabilities.  By
considering surface wave motions as a perturbation to the underlying mean
flows, we treat the surface waves as being reflected or refracted much
like wave scattering in optical, acoustic, or electronic solid state
physics, where band structures have been calculated using linear
eigenvalue analysis in various systems \cite{ASHCROFT}. A related problem
of periodic surface wave scatterers such as thin ice floes has been
treated with similar methods \cite{CHOU97}.  Here the underlying flow is
periodic so we use Bloch functions to describe the surface displacements
and dynamic fluid pressure and derive a generalized quadratic eigenvalue
equation with a nonhermitian operator.  In addition to nontrivial ``band
structure'' we find a complex eigenvalue spectrum corresponding to linear
instabilities with rich behavior. We show the existence of interfacial
instabilities in the one layer problem, as well as modifications to the
stability regions of uniform shear in the two-layer problem (K-H
instability) when an additional periodic flow is present.

\begin{figure}[htb]
\begin{center}
\leavevmode
\epsfysize=1.5in
\epsfbox{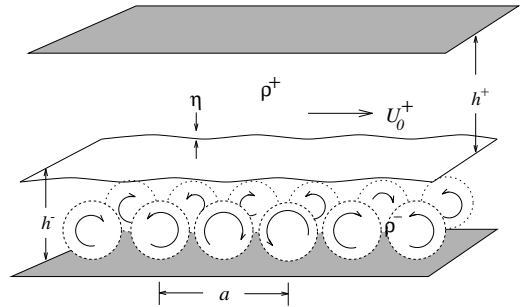}
\end{center}
\caption{Schematic of problem geometry. Two ideal fluids are separated by
an interface with surface tension $\sigma$ under the action of gravity. 
The densities, velocities, and depths of the upper and lower fluids
are $\rho^{\pm}$, $\vec{U}^{\pm}$, and $h^{\pm}$ respectively.}
\label{Fig1}
\end{figure}

%\begin{figure}[htb]
%\begin{center}
%\leavevmode
%\epsfysize=1.5in
%\epsfbox{Fig1.eps}
%\end{center}
%\caption{Schematic of problem geometry. Two ideal fluids are separated by
%an interface with surface tension $\sigma$ under the action of gravity. 
%The densities, velocities, and depths of the upper and lower fluids
%are $\rho^{\pm}$, $\vec{U}^{\pm}$, and $h^{\pm}$ respectively.}
%\label{FIG1}
%\end{figure}

%The flows corresponding to these flow patterns can affect
%the propagation of surface waves. The flows alter the
%surface dynamical boundary condition which in turn
%determine the surface wave dispersion relation.  Waves can
%therefore refract or reflect from regions of surface
%shear. Flow-wave interaction will clearly be relevant when
%the flow velocity is a nonnegligible fraction of the 
%wave group velocity. For the example of Rayleigh-B\'{e}nard
%convection, the typical velocity of the convection 
%rolls scales as $...$.

{\it Formulation - } Consider two ideal fluids separated by a flat
wave-undisturbed interface at $z=0$ as depicted in Fig. 1.  Although the
underlying flow field is typically rotational, we assume the flow
associated with the imposed surface waves (externally excited)
are irrotational.  The displacements caused by the surface
waves are assumed small and do not appreciably affect the underlying
flow. This requirement, and linearization, implies $\eta \ll \lambda, a$
where $\eta, \,\lambda$, and $a$ are typical surface amplitudes,
wavelengths, and flow field periodicities respectively.

%The boundary at $z=-h$ is taken to be impenetrable
%such that $\partial_{z}\phi(z=-h) = 0$.  The boundary at $z=-h$
%is taken to be impenetrable such that $\partial_{z}\phi(z=-h) =
%0$.  The boundary at $z=-h$ is taken to be impenetrable such that
%$\partial_{z}\phi(z=-h) = 0$.  
 
The fluid velocities above and below the interface are $\vec{v}^{\pm} =
\vec{U}^{\pm} + \nabla\varphi^{\pm}$ where $\vec{U}$ is the periodic flow
field (generated for example, by the Rayleigh-B\'{e}nard
instabilities\cite{LEAL}, Langmuir circulation mechanisms\cite{LC}, or
electrohydrodynamic effects\cite{AJDARI}) which satisfies
$\nabla\!\cdot\!\vec{U}=0$, and $\varphi$ is the velocity potential for
the irrotational capillary-gravity waves, and $z= \pm h^{\pm}$ is the
position of the impenetrable top and bottom.  For concreteness, consider
the Rayleigh-B\'{e}nard system where instabilities to periodic flows in
the Boussinesq approximation with a free surface arises when Ra $\equiv
\alpha g h^{3} \Delta T/(\nu\kappa) > Ra^{*}$, (Ra$^{*} \simeq 1100$ for
a free, tensionless surface) where $\alpha,\,g,\,\Delta T,\, \nu,\,$ and
$\kappa$ are the thermal expansion coefficient, gravitational
acceleration, temperature difference $T(z=0)-T(z=-h)$, kinematic
viscosity, and thermal conductivity, respectively. The ideal fluid
approximation is valid only for surface waves which are not significantly
damped over many periods of the underlying flow, {\it i.e.,} for
capillary waves, the attenuation length $k_{d}^{-1} \equiv 3\sigma
/(4\nu\rho^{-}\omega) \gg a$. We also assume in this system a large
Prandtl number such that the Reynolds number, Re $\sim
\sqrt{\mbox{Ra/Pr}}$ is not too large as to render the steady periodic
flows unstable.  

%Furthermore, we require, as an approximate condition for stability of
%the periodic flows, that the associated Reynolds number, Re $\equiv
%Ua/\nu \lesssim 1$.  Using a typical critical velocity, $U_{c} \sim
%$(\alpha g h \Delta T)^{1/2}$, we find Re $\sim \sqrt{\mbox{Ra/Pr}}$,
%$where Pr $\equiv \nu/\kappa$ is the Prandtl number.  Therefore, all
%$conditions required for our assumptions are met for Pr $>$ Ra$^{*}\simeq
%$1100$, or $\kappa \lesssim 6.82\times 10^{-4}\sigma/(\rho \omega a)$
 
Incompressibility demands $\nabla\!\cdot\!\vec{v} = \Delta \varphi =
(\Delta_{\perp} + \partial_{z}^{2}) \varphi^{\pm} = 0$, where
$\Delta \equiv \Delta_{\perp} + \partial_{z}^{2}$ is the three
dimensional Laplacian.  The boundaries at $z=\pm h^{\pm}$ are
assumed flat and impenetrable.  Therefore, the linearized kinematic
boundary conditions at $z=\pm h^{\pm}$ and the interface at $z
\simeq 0$ are $\partial_{z}\varphi^{\pm}(\vec{x},z=\mp h) = 0$ and

\begin{equation}
\partial_{t}\eta(\vec{r})+\vec{U}^{\pm}(\vec{r})\!\cdot\!
\nabla_{\perp}
\eta(\vec{r}) = \lim_{z\rightarrow 0^{\pm}}\,
\partial_{z}\varphi^{\pm}(\vec{r},z)
\label{KINBC1}
\end{equation}

\noindent respectively, where $\vec{r}_{\perp}\equiv \vec{r}
 \equiv (x, y)$.  The linearized dynamical
boundary condition at $z\simeq 0$ is found by balancing 
$z-$component stresses from the dynamical pressure with that
from gravity and surface tension;

\begin{equation}
\lim_{z\rightarrow 0^{\pm}}\left[\partial_{t}\varphi^{\pm} + 
U^{\pm}\partial_{i}\varphi^{\pm}\right] = 
{\sigma \over \rho^{\pm}}\Delta_{\perp}\eta - g\eta,
\label{BC}
\end{equation}

\noindent where we have for simplicity assumed a constant surface tension
and neglected any possible Marangoni effects that may arise when
surfactants are convected along the surface by $\vec{U}(\vec{r}, z=0)$. 
Spatially varying surface properties such as tension or bending rigidity
can be treated without difficulty \cite{CHOU97,CHOU94}.  Henceforth, we
assume that all quantities vary as $e^{-i\omega t}$. Wave evolution due
to a non-time harmonic source can be found by superposing the solutions
of many frequency components. Consider general solutions to
$\varphi^{\pm}$ and $\eta(\vec{x})$, 

\begin{equation}
\varphi^{\pm}(\vec{r},z) =
\sum_{\vec{q}}\varphi^{\pm}_{\vec{q}}e^{i\vec{q}\cdot\vec{r}}
{\cosh q(h^{\pm}\mp z) \over \cosh qh^{\pm}},
\label{PHIQ}
\end{equation}

\noindent and

\begin{equation}
\eta(\vec{r}) =
\sum_{\vec{q}}\eta_{\vec{q}}e^{i\vec{q}\cdot\vec{r}}
\label{ETAQ}
\end{equation}

\noindent where $\vec{q} \equiv \vec{q}_{\perp}$ lies in the surface
plane, and $q \equiv \lim_{\epsilon\rightarrow
0}\sqrt{q^2+\epsilon^{2}}$. Equation (\ref{PHIQ}) automatically
satisfies $\nabla^{2}\varphi = 0$ as well as the impenetrable bottom
boundary condition. Therefore by using Eqn. (\ref{PHIQ}), the
problem is reduced to that of simultaneously solving equations
(\ref{KINBC1}) and (\ref{BC}) with the unknowns
$\varphi^{\pm}_{\vec{q}}$ and $\eta_{\vec{q}}$. A Fourier
decomposition for the periodic flows is

\begin{equation}
\vec{U}(\vec{r},z) =
\sum_{\vec{G}}\vec{U}(\vec{G},z)e^{i\vec{G}\cdot\vec{r}},
\label{UDECOMP}
\end{equation}

\noindent where $\vec{G}$ is the reciprocal lattice vectors
appropriate for the underlying flow periodicity.  With the form
(\ref{UDECOMP}), the velocity potential $\varphi$ will furthermore
be restricted to forms satisfying Bloch's theorem\cite{ASHCROFT}:
$\varphi(\vec{r}+\vec{a}, z) = e^{i\vec{q}\cdot\vec{r}}f(\vec{r},z)$ where
$f(\vec{r})$ is a function periodic with respect to translations of
$\vec{a}$.

We wish to find the complex dispersion
relation between $\omega$ and wavevector $\vec{q}$.  
Substituting Eqns. (\ref{PHIQ}) and (\ref{ETAQ}) into 
(\ref{KINBC1}) and (\ref{BC}), we obtain 

\begin{equation}
\varphi^{\pm}_{\vec{q}} = \pm {1 \over q\tanh
qh^{\pm}}\left[ i\omega\eta_{\vec{q}} - i\sum_{\vec{G}}
\vec{U}^{\pm}(\vec{G})\cdot\!(\vec{q}-\vec{G})
\eta_{\vec{q}-\vec{G}}\right]
\label{PHIQ2}
\end{equation}

\noindent and 

\begin{equation}
\begin{array}{l}
\displaystyle s\left[-i\omega\varphi^{+}_{\vec{q}}+i\sum_{\vec{G}}
\vec{U}^{+}(\vec{G})\!\cdot\!(\vec{q}-\vec{G})
\varphi^{+}_{\vec{q}-\vec{G}}\right] +  \\[13pt]
\displaystyle \quad\quad \left[i\omega\varphi^{-}_{\vec{q}} -
i\sum_{\vec{G}}\vec{U}^{-}(\vec{G})\!\cdot\!(\vec{q}-\vec{G})
\varphi^{-}_{\vec{q}-\vec{G}}\right] = 
\left({\sigma \over \rho}q^2 + g\right)\eta_{\vec{q}}
\label{BC2}
\end{array}
\end{equation}

\noindent respectively. Shifting the the wavevector into the first
Brillouin zone, and substituting (\ref{PHIQ2}) into (\ref{BC2}), we
find a generalized eigenvalue problem

\begin{equation}
\sum_{\vec{G}'}\left({\bf A}\omega^{2}+{\bf B}\omega + {\bf C}\right)
\eta_{\vec{q}}(\vec{G}') = 0
\label{MATRIX0}
\end{equation}

\noindent where the matrices in the limit of 
unbounded upper fluid ($h^{+} = \infty$, $h\equiv h^{-}$)
are given by

\begin{equation}
A(\vec{G}, \vec{G}') \equiv (s\tanh \vert\vec{q}-\vec{G}\vert h +
1)\delta_{\vec{G},\vec{G}'}
\end{equation}

\begin{equation}
\begin{array}{l}
\displaystyle B(\vec{G}, \vec{G}') \equiv - 2s\,
\vec{U}^{+}_{0}\!\cdot\!(\vec{q}-\vec{G})
\tanh\vert\vec{q}-\vec{G}\vert h\,
\delta_{\vec{G},\vec{G}'} \\[13pt]
\displaystyle \quad - \left(
1+{\vert \vec{q}-\vec{G}\vert\tanh\vert \vec{q}-\vec{G}\vert h 
\over \vert \vec{q}-\vec{G}'\vert \tanh\vert
\vec{q}-\vec{G}'\vert h}\right)
\vec{U}^{-}(\vec{G}-\vec{G}')\!\cdot\!(\vec{q}-\vec{G}')
\end{array}
\end{equation}

\begin{equation}
\begin{array}{l}
\displaystyle \!C(\vec{G},\vec{G}') \equiv \left(s 
\vert\vec{U}^{+}_{0}
\!\cdot\!(\vec{q}-\vec{G})\vert^{2}\tanh\vert
\vec{q}-\vec{G}\vert h -\Omega_{\vec{q}}^{2}(\vec{G})
\right) \delta_{\vec{G},\vec{G}'}\\[13pt]
\displaystyle  + \sum_{\vec{G}''}
{\vert\vec{q}-\vec{G}\vert\tanh\vert\vec{q}-\vec{G}\vert h \over
\vert\vec{q} - \vec{G}''\vert\tanh\vert\vec{q}-\vec{G}''\vert h}
\vec{U}^{-}(\vec{G}-\vec{G}'')\!\cdot\!(\vec{q}-\vec{G}'') \\[13pt]
\quad\mbox{\hspace{3cm}}\times\vec{U}^{-}(\vec{G}''-\vec{G}')\!\cdot\!(\vec{q}-\vec{G}')
\end{array}
\label{MATRIXC}
\end{equation}

\noindent where $s \equiv \rho^{+}/\rho^{-}$, and 

\begin{equation}
\Omega^{2}_{\vec{q}}(\vec{G}) = \left({\sigma \over \rho^{-}}\vert
\vec{q}-\vec{G}\vert^{3} + (1-s)g
\vert \vec{q}-\vec{G}\vert\right)\tanh\vert\vec{q}-\vec{G}\vert h
\end{equation}

\noindent We arrive at an eigenvalue problem that can be solved by
standard means \cite{RECIPES},

\begin{equation}
\left(
\begin{array}{cc} {\bf 0} & {\bf 1} \\
-{\bf A}^{-1}{\bf C}\quad & -{\bf A}^{-1}{\bf B}
\end{array}
 \right)\left(\begin{array}{l} \vec{\eta}_{\vec{q}}\\
\vec{\psi}_{\vec{q}} \end{array} \right)
 = \omega \left(\begin{array}{l} \vec{\eta}_{\vec{q}} \\
\vec{\psi}_{\vec{q}} \end{array} \right),
\label{MATRIX}
\end{equation}

\noindent where $\vec{\psi} \equiv \omega\vec{\eta}$. However, the
above matrix is generally nonhermitian and the corresponding spectrum
$\omega$ is complex. The standard criteria for the Kelvin-Helmholtz
propagating wave instability is recovered when $\vec{U}^{\pm} =
\vec{U}_{0}^{\pm}$ are uniform\cite{WHITHAM}:

\begin{equation}
(\omega-\vec{U}_{0}^{-}\!\cdot\!\vec{k})^2+s\tanh kh\,(\omega -
\vec{U}_{0}^{+}\!\cdot\!\vec{k})^2 - \Omega_{k}^{2}(0) = 0.
\label{KHUNIFORM}
\end{equation}

\noindent Note that when dynamical effects of the upper fluid
are neglected, ($s = 0$), the roots of Eqn. 
(\ref{KHUNIFORM}) are real, and no linear instability exists.  

{\it Results and Discussion -} Distances, frequencies $\omega$, and
velocities $U$ will be normalized and measured in units of $a$,
$\sqrt{g/a}$, and $\sqrt{ga}$ respectively. First consider only
periodic flow in the lower fluid with no uniform component,
$\vec{U}^{\pm}(\vec{G}=0) = 0$. Only the velocity at the surface
$\vec{U}^{-}(\vec{r}, z=0)$ will influence surface {\it wave}
propagation.  For simplicity we only analyze one-dimensional rolls
described by the approximate function,

\begin{equation}
\vec{U}^{-}(\vec{r}, z=0) = U^{-}(z)\hat{x}\cos
\left({2\pi \over a}x\right);
\end{equation}

\noindent consequently, $\vec{U}^{-}(\vec{G}) = U^{-}(0)\hat{x}/ 2\pi$ for
$\vec{G} = \pm 2\pi\hat{x}/a$, and zero otherwise.  This choice of phase for
$\vec{U}^{-}(\vec{r}, z=0)$ implies one converging and one diverging surface
flow region per unit cell and also simplifies the computation by making all
elements of the matrix in Eq. (\ref{MATRIX}) real. Also, we choose small
surface tensions so that the periodic flow surface velocities will be
nonnegligible compared to water wave group velocities.  

\begin{figure}[htb]
\begin{center}
\leavevmode
\epsfysize=3.5in
\epsfbox{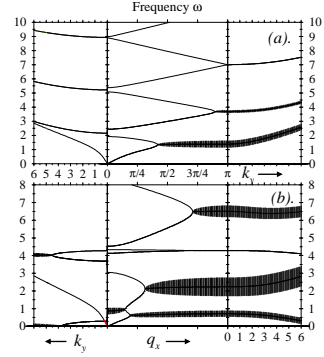}
\end{center}
\caption{Triptych depicting the band structure for periodic flow
in the lower fluid ($s=0,\,h=2.0,$ Bo$^{-1}=0.01$). The central panel shows the
dispersion relation in the $q_{x}$ direction in the reduced zone scheme.  (a).
$U^{-}(2\pi/a) = 0.20$; (b). $U^{-}(2\pi/a) = 0.75$.}
\label{Fig2}
\end{figure}

We find the spectrum of (\ref{MATRIX}) using standard methods
\cite{RECIPES} and thus obtain the band structure of capillary-gravity
waves on surfaces with underlying periodic flows $\vec{U}^{-}$.  The
central panels in Figures \ref{Fig2} show the real (solid lines) and
imaginary (height of hatched regions centered about Re$\{\omega\}$) parts
of $\omega(q_{x},0)$. We use the notation $q\,(k)$ to denote quantities
plotted in the reduced(extended) zone scheme. The side panels show
$\omega(0,k_{y})$ for $\vec{U}^{\pm}_{0}=0$, $h=2.0$, $s=0$, and the
inverse Bond number Bo$^{-1} \equiv \sigma/(\rho^{-}a^2 g) = 0.01$. 
$U^{-}(\pm 2\pi/a) = 0.2,\, 0.75$ are shown in (a). and (b). 
respectively.  The ``band structures'' shown in 2(a). contain branch cuts
at certain $q_{x}$ satisfying the Bragg scattering condition. There are
open band gaps at $q_{x}=0$, which decrease at larger $\omega$, similar
to electronic and acoustic wave propagation in periodic media; the gaps
normally found at $q_{x}=\pm \pi$ are collapsed due to the converging
flow in each unit cell and are degenerate down to a smaller value of
$q_{x}$. Under this periodic flow, growing surface modes arise near
$q_{x}\sim\pm \pi$ when real degenerate roots of Eqn. (\ref{MATRIX})
split into complex conjugate pairs, even though for the one fluid
problem, {\it uniform stream} flows are stable according to
(\ref{KHUNIFORM}).  In Fig \ref{Fig2}(b), $\vert U^{-}(\pm 2\pi/a) \vert
= 0.75$ where the first gap at $q_{x}=0$ has also merged. Note that 
unstable modes are associated with both standing and low group velocity
travelling waves since they appear predominantly near $q_{x}=0, \pi$.
These may be damped or saturated by viscous dissipation, Im$\{\omega\}
\simeq 2\nu \vert \vec{q}-\vec{G}\vert^{2}$. For
$\vec{U}^{-}(2\pi/a)<0.3641$, or higher values of surface tension, the
band structure qualitatively resembles that of Fig. 2(a). At intermediate
values of $\vec{U}^{-}(2\pi/a)$ (not shown), we find unstable
nonoscillatory modes; {\it i.e.} $Re\{\omega(q_{x},k_{y})\}=0,\,
Im\{\omega(q_{x},k_{y})\}<0$. When $0.417>\vec{U}^{-}(2\pi/a)>0.3641$,
the lowest two branches collapse such that unstable zero frequency modes
proliferate and fill the whole zone.  These growing modes will be
temporally saturated by viscosity and/or nonlinear effects, and result in
static surface deformations similar to the Reynolds ridge \cite{STATIC}.
Upon further increasing $\vec{U}^{-}(2\pi/a) > 0.417$, the zero frequency
growing modes develop a finite frequency and stabilize near $q_{x} \sim
0$. When $\vec{U}^{-}(2\pi/a) \simeq 0.428$, the first gap at $q_{x}=0$
between the stable modes collapse. The qualitative behavior described
above continually repeats upon increasing $\vec{U}^{-}(2\pi/a)$.  Note
that for higher $U^{-}(2\pi/a)$ there are specific directions where
oscillatory and nonoscillatory ($Re\{\omega(q_{x},k_{y})\} \rightarrow
0$) instabilities also arise.

\begin{figure}[htb]
\begin{center}
\leavevmode
\epsfysize=2.1in
\epsfbox{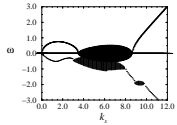}
\end{center}
\caption{Dispersion relation with and without an underlying
periodic flow with a uniform ($\vec{U}_{0}^{+}=2.0$) overlying
shear.  Here, $h=\infty$, Bo$^{-1} = 0.03$, and $s=0.1$. As a
visual aid, we have imposed a $k_{x}$ dependent Galilean shift
$\Delta \vec{U}_{0} = (\vec{U}_{0}^{-}+\vec{U}_{0}^{+}s\tanh
k_{x}h)/(1+s\tanh k_{x}h)$.  The pure shear dispersion
($\vec{U}^{-} = 0$) is shown by the dotted lines, with the
imaginary parts of the unstable growing modes depicted by the
heights of the hatched regions. The solid line ($\omega<0$)
defines the dispersion relation when a periodic flow
($\vec{U}^{-}(\vec{G}\neq 0)= 0.2$) is imposed. Note the changes
to the instability regimes. }
\label{Fig3}
\end{figure}

The band structures shown in Figures 2 are only quantitatively altered
(tilted) when a small uniform shear is imposed in addition to the
periodic flow.  However, for uniform shear ($s = 0.1, h=\infty, U_{0}^{+} = 2.6,$
Bo$^{-1}$ = 0.1) flows with an existing K-H instability
(with unstable wavevectors near $k_{x} \sim \pi$) as determined by
Eq. (\ref{KHUNIFORM}), the effect of an additional periodic flow in the
lower fluid enhances these instabilities.  However, when the
uniform shear instabilities span $k_{x} \sim 2\pi$ ($s = 0.1, h=\infty, U_{0}^{+} =
2.0,$ Bo$^{-1}$ = 0.03) where there is an open gap, the K-H instability
can be {\it suppressed} by a periodic flow $U^{-}(2\pi/a) = 0.20$,
although more instabilities arise at higher $k_{x}$ as well. These
effects are shown using the extended zone scheme in Figure 3. The
structure of the dispersion relation is rather sensitive to the amount of
underlying periodic flow and can change drastically with variation in any
of the parameters.  

Qualitatively, the upward convex tongue of the K-H instability in the
$k_{x}-U_{0}^{+}$ plane is modified when $U^{-}(\vec{G}\neq 0)$ is added.
For instabilities straddling the open gap Bragg planes, this tongue is
shifted, with parts shifted upwards (resulting in destabilization) and
downwards(resulting in stabilization). The discontinuity in growth rate
is clearly shown in Fig. 3 near $k_{x} \simeq 2\pi$.  Moreover, other
smaller tongues develop at higher $q_{x}$ ($k_{x}\simeq 3\pi$ in Fig. 3)
corresponding to the closed gap instabilities that would occur in
exclusively periodic flows (Fig. 2). The growth rates within the
instability regions change as well as the delimiting instability regions.
Qualitatively, these effects on K-H instabilities result from suppression
of the travelling unstable waves in K-H shear flow, due to coupling to
the standing waves of periodic flow, arising in the last term in ${\bf
C}$ in Eq. (\ref{MATRIXC}).
 
We have shown that periodic flows couple different reciprocal wavevectors
together and affect interfacial stability in a nontrivial manner. In
particular, by solving the matrix equations, we find, remarkably, that an
interface overlying periodic flows is generally linearly unstable, {\it
even if} $\rho^{+} = 0$.  The linear instabilities described would be
saturated by both viscous and nonlinear effects.  Regions of instability
corresponding to finite frequency modes with vanishing group velocities,
dissipation due to viscosity will result in a standing oscillating mode.

The analogies between wave propagation in periodic media and surface wave
propagation in periodic shear flow configurations can be extended to
consider more complicated periodic flow structures such as rectangular
and hexagonal patterns, where rich behavior should be expected. This may
lead to insights with wide applicability, from Rayleigh-B\'{e}nard
convection, Langmuir circulation, solar convection cells, and MHD surface
Alfv\'{e}n waves in the presence of periodic magnetic fields (in addition to
flows) \cite{SHERCLIFF}. Furthermore, the influence of defects and
disorder in the periodic surface flows\cite{DISORDER} can be considered
to study surface wave localization in the presence of random
$\vec{U}(\vec{r},0)$ (and hence random ${\bf B},\,{\bf C}$). For example,
methods used to determine the complex spectrum density of states of
random matrix operators\cite{WANG} should yield interesting behavior
regarding the sea surface wave spectra in the presence of random
underlying turbulent surface flows. 

\vspace{2mm}

%\noindent The author is grateful for support from The Wellcome
%Trust.  

\end{document}